# Electrically Tunable Room Temperature Hysteresis Crossover in Underlap MoS₂ FETs

Himani Jawa,[†] Abin Varghese,[†,‡] and Saurabh Lodha[*,†]

†Department of Electrical Engineering, IIT Bombay, Mumbai, Maharashtra 400076

‡Department of Materials Science and Engineering, Monash University, Clayton, Victoria, Australia 3800

E-mail: slodha@ee.iitb.ac.in

## Abstract

Clockwise to anti-clockwise hysteresis crossover in current-voltage transfer characteristics of field effect transistors (FETs) with graphene and MoS₂ channels holds significant promise for non-volatile memory applications. However such crossovers have been shown to manifest only at high temperature. In this work, for the first time, we demonstrate room temperature hysteresis crossover in few-layer MoS₂ FETs by using a gate-drain underlap design to induce a differential response from traps at the MoS₂-HfO₂ channel-gate dielectric interface to applied gate bias. The appearance of interface trap-driven anti-clockwise hysteresis at high gate voltages in underlap FETs can be unambiguously attributed to the presence of an underlap since transistors with and without the underlap region were fabricated on the same MoS₂ channel flake. The underlap design also enables room temperature tuning of the anti-clockwise hysteresis window (by 140×) as well as the crossover gate voltage (by 2.6×) with applied drain bias and underlap length. Comprehensive measurements of the transfer curves in ambient and vacuum conditions at varying sweep rates and temperatures (RT, 45 ℃ and 65 ℃) help segregate the quantitative contributions of adsorbates, interface traps, and bulk HfO₂ traps to the clockwise and anti-clockwise hysteresis.

# Introduction

Recently, van der Waals (vdW) materials have attracted extensive research interest due to excellent electronic, optical and mechanical properties arising from their two-dimensional (2D) and layered nature. Graphene, the first 2D material discovered in 2004, suffers from a lack of bandgap, making semiconducting transition metal dichalcogenides (TMDs) with a bandgap[1–3] more promising for electronic applications. Among the TMDs, $MoS_2$ with a direct bandgap of 1.9 eV in monolayer form and an indirect bandgap of ~1.2 eV in bulk has emerged as a popular choice.[4,5] Large electron mobilities, high on-current/off-current ($I_{ON}/I_{OFF}$) ratios and low off-currents have been demonstrated on ultra-thin $MoS_2$ based field effect transistors (FETs).[6] In addition, $MoS_2$ based devices have been employed in gas/chemical sensing, neuromorphic, memory, optoelectronic and RF switching based applications.[7–19]

Instability[20–22] and hysteresis[23–27] in the threshold voltage ($V_T$) of $MoS_2$ transistors are serious bottlenecks in their use for various applications. Although $V_T$ hysteresis is a concern for most applications, a few studies have shown how it can be positively harnessed for thermally-assisted memories. Specifically, hysteresis crossover (anti-clockwise (ACW) to clockwise (CW) in a given gate voltage double sweep at high temperature) and hysteresis inversion (CW for room temperature gate voltage double sweep to ACW for higher temperature gate voltage double sweep) have been demonstrated. However, such behaviour in hysteresis characteristics has been obtained at temperatures >350 K for $MoS_2$ transistors hysteresis crossover at ~400 K for a monolayer $MoS_2$ FET[17] and hysteresis inversion at >350 K for bulk $MoS_2$ FETs.[18] Besides low temperature behavior, electrical tunability of hysteresis crossover has not been demonstrated yet. Further, multiple sources for $V_T$ hysteresis[24–27] reported in previous studies include traps introduced by adsorbates[28,29] (ambient water and oxygen molecules), interface traps at the $MoS_2$-gate dielectric interface[30–34] and traps near the gate metal-gate dielectric interface.[17] The contribution of each type of trap to $V_T$ shift and hysteresis varies with the device structure, ambient conditions, measurement temperature, applied gate voltage and its sweep rate and direction. A systematic delineation and quantitative estimate of these contributions are essential for addressing them.

This work demonstrates electrically tunable room temperature (RT) hysteresis crossover (from CW at lower gate voltages to ACW at higher gate voltages) as well as quantitative contributions of various kinds of traps, viz. adsorbates, channel-gate dielectric interface traps and traps in the gate dielectric, through comprehensive electrical measurements with varying temperature and ambient conditions in a gate underlap few-layer $MoS_2$ FET. Fabrication and characterization of fixed channel length (3 $\mu$m), gate overlap and underlap (200 nm) FETs on the same $MoS_2$ flake, to circumvent flake-to-flake variability, shows the emergence of CW-to ACW hysteresis crossover at RT with the gate underlap. Primarily, this gate-drain underlap engineers a differential response from channel-gate dielectric interface traps in the overlap and underlap gate regions, of the same transistor and at the same gate voltage, through a dual- the direct (overlap) and the fringing (underlap)- gate capacitance effect. A gate underlap $MoS_2$ FET design can hence enable RT memory applications. The crossover (CW to-ACW) gate voltage can be tuned with drain voltage as well as the underlap length, and is mediated by traps at the $MoS_2/HfO_2$ interface in the underlap region. A 2.6× variation in crossover gate voltage and 140× modulation of the anticlockwise hysteresis window for drain bias ranging from 0.1-0.4 V can enable tunability of read and write memory parameters.[17,18] Analysis of the hysteresis widths, relative $V_T$ shifts and subthreshold slopes (SS) of the two transistors under ambient and vacuum conditions and with varying temperature (RT, 45 ℃ and 65 ℃) helps extract the densities of adsorbates, traps at the $MoS_2/HfO_2$ interface and traps in the $HfO_2$ dielectric.

## Results and discussion

Figure 1a shows a 3D schematic of the bottom gate underlap and overlap $MoS_2$ FETs used in this work. The gates were first patterned on an $Si/SiO_2$ substrate using electron beam lithography (EBL) and metal deposition (Cr/Au ∼ 10/80 nm). This was followed by gate dielectric deposition (16 nm $HfO_2$) using atomic layer deposition (ALD) at 200 ℃. $MoS_2$ flakes were mechanically exfoliated from an $MoS_2$ crystal onto a polydimethylsiloxane (PDMS) stamp and a few-layer flake was transferred onto the patterned gates using a dry transfer process. Thickness of the $MoS_2$ flake was measured to be ∼7 nm (10 layers) using atomic force microscopy (AFM), as shown in Fig. 1b. Further, source/drain contacts were formed using EBL and metal deposition (Cr/Au ∼ 10/80 nm) ensuring a gate-drain separation of 200 nm (0 nm) in the underlap (overlap) FET. Finally, the HfO$_2$ layer over the gate contact pads was etched using buffered HF (BHF) solution. Optical micrograph of the FETs is shown in Fig. 1c, where multiples devices were made on the same MoS$_2$ flake with varying gate underlap lengths, keeping the channel length constant (3 $\mu$m) for all FETs. However, the data discussed in this work was primarily measured on two FETs- gate G$_1$, drain D$_1$ and source S$_1$ of the overlap device (gate length=channel length= 3 $\mu$m) and gate G$_2$, drain D$_2$ and source S$_2$ of the underlap FET (gate length= 2.8 $\mu$m, channel length= 3 $\mu$m). Figure 1d(i) shows a scanning electron microscope (SEM) image of both the overlap and the underlap devices. Figure 1d(ii) shows the magnified SEM image of the region between D$_2$ and G$_2$ indicating an underlap of ~200 nm.

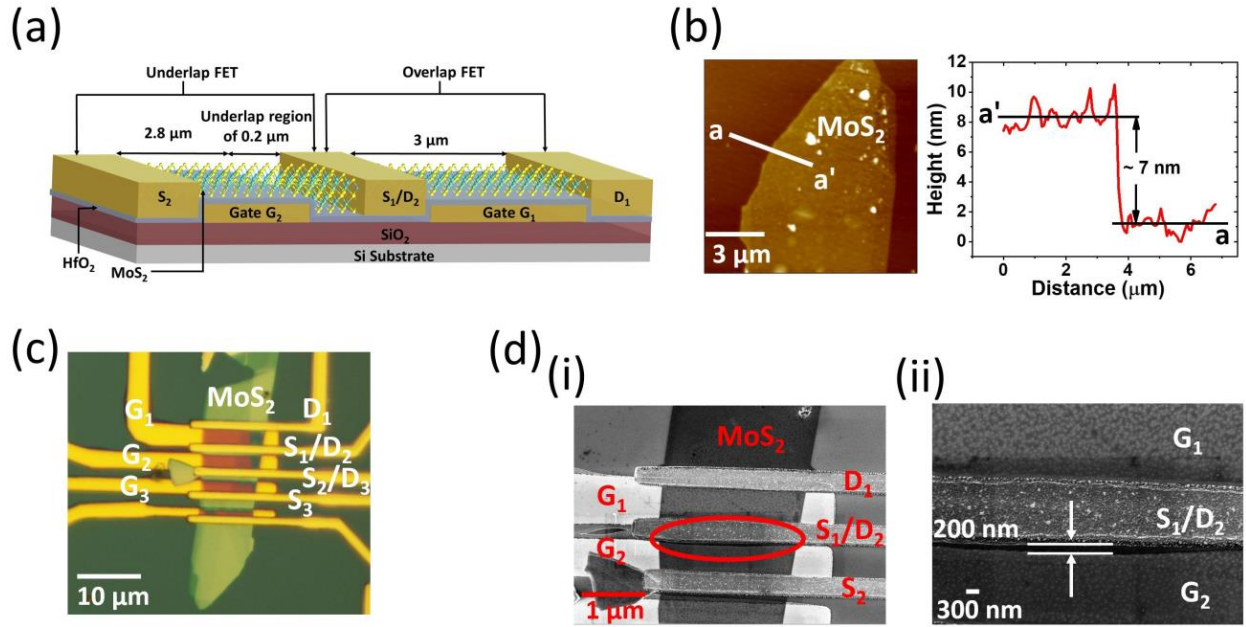

Figure 1: (a) 3D schematic of underlap and overlap FETs with MoS$_2$ as the channel material and HfO$_2$ as the bottom gate dielectric. (b) AFM scan confirms a 7 nm thick (~10 layers) MoS$_2$ flake. (c) Optical microscope image of the devices with gates (G$_1$ and G$_2$) at the bottom and source/drain contacts (D$_1$, S$_1$ and D$_2$, S$_2$) on the top. G$_1$, D$_1$ and S$_1$ form the overlap device and G$_2$, D$_2$ and S$_2$ form the underlap device. (d) SEM images showing both overlap and underlap FETs in (i) with a 200 nm underlap separation between D$_2$ and G$_2$, enlarged and shown in (ii).

Figure 2a shows the double sweep (forward sweep, FS and reverse sweep, RS) transfer characteristics (I$_D$-V$_G$) of the underlap as well as the overlap device for an applied drain voltage (V$_D$) of 0.2 V at RT in ambient conditions. Overlap FET demonstrates clockwise hyste-resis for the entire $V_G$ range whereas clockwise hysteresis at low $V_G$ crosses over to anti-clockwise hysteresis at high $V_G$ for the underlap FET. Figure 2b shows that clockwise hysteresis in the overlap device is seen for a wide range of $V_D$ as it is varied from 0.1 V to 0.7 V. Figure 2c shows the transfer characteristics for the underlap device with varying $V_D$ (0.1 to 0.7 V). $\Delta V_T$, calculated as the difference between RS ($V_G$ from +3 to -2 V) and FS ($V_G$ from -2 to +3 V) threshold voltages ($V_T$) at a constant current of ~1e-7 A, shows almost no dependency on drain voltage and hence, the clockwise hysteresis is similar for both devices, as shown in the insets of Fig. 2b and 2c for the overlap and underlap FETs, respectively. However, to describe the effect of $V_D$ on anti-clockwise hysteresis for the underlap device, two parameters have been defined: firstly, $V_{GC}$, the $V_G$ at which clockwise hysteresis crosses over to anti-clockwise hysteresis for given $V_D$, and secondly, $\Delta I_D/I_{D,FS}$, the difference in current ($\Delta I_D$) between the RS and FS drain currents ($I_{D,RS}$ and $I_{D,FS}$ respectively) normalised to the FS drain current at a constant $V_G$ (>$V_{GC}$). As shown in the inset of Fig. 2c, $\Delta I_D/I_{D,FS}$ (extracted at $V_G$ = 2 V) reduces by 140× as $V_D$ is increased from 0.1-0.4 V indicating a reduction in anti-clockwise hysteresis for a given gate voltage. This is also supported by a 3.64× (2.6×) increase in $V_{GC}$ with $V_D$ increasing from 0.1-0.7 V (0.1-0.4 V), as shown in Fig. 2d. Inset of Fig. 2d shows that $V_{GC}$ increases, anti-clockwise hysteresis decreases and the gate control ($I_{ON}/I_{OFF}$) worsens when the underlap length ($L_U$) is increased from 200 to 700 nm.

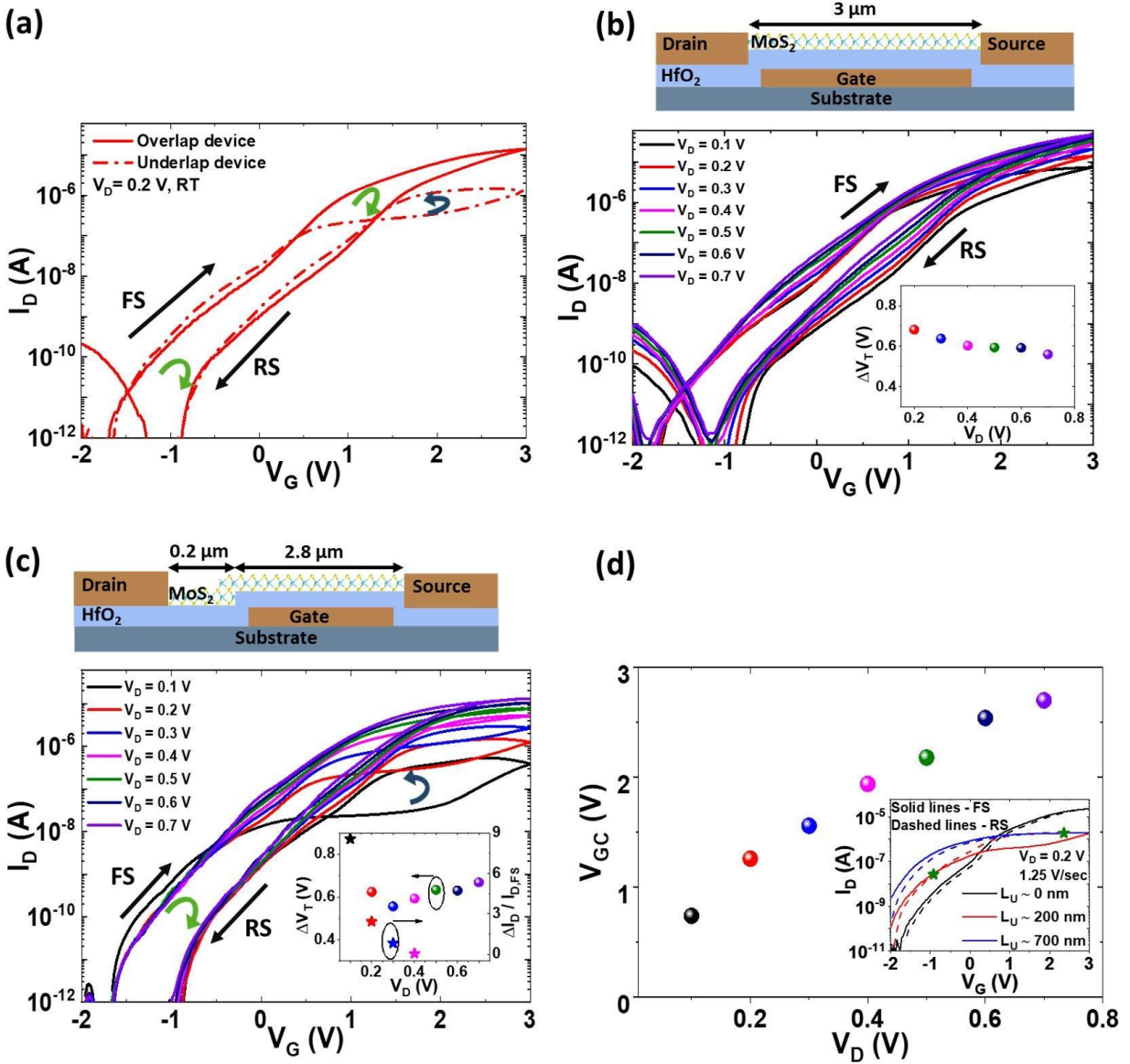

Figure 2: (a) $I_D$-$V_G$ transfer characteristics at $V_D$ = 0.2 V for the underlap and overlap devices at RT in ambient. The overlap device shows only clockwise hysteresis whereas the underlap device shows clockwise hysteresis at lower $V_G$ that crosses over to anti-clockwise hysteresis at higher $V_G$. FS and RS represent forward and reverse sweep, respectively. $I_D$-$V_G$ of the (b) overlap and (c) underlap FETs for varying $V_D$ with their respective 2D schematics. The inset in (b) shows $\Delta V_T$ staying nearly constant with varying drain voltage whereas the inset of (c) shows both $\Delta V_T$ and $\Delta I_D/I_{D,FS}$ (extracted at $V_G$ = 2 V) for varying drain voltage. $\Delta V_T$ being constant for both devices indicates $V_D$-independent clockwise hysteresis. (d) Crossover voltage ($V_{GC}$) versus drain voltage for underlap FET. Inset shows the transfer characteristics for varying underlap length (0, 200 and 700 nm). The crossover point is absent in the overlap device (D1-S1) and $V_{GC}$ increases with increasing underlap length (D2-S2, $L_U \sim$200 nm and D3-S3, $L_U \sim$700 nm).

Clockwise hysteresis present in the underlap and overlap FETs can be explained by $MoS_2/HfO_2$ interface traps and adsorbate mediated mechanisms.[28,29,31,33] However, hysteresis crossover (CW to ACW), present only in the underlap device, is primarily due to the action of $MoS_2/HfO_2$ interface traps. For the n-channel $MoS_2$ overlap device, electron trapping at the $MoS_2/HfO_2$ interface, and at adsorbate sites, is suppressed at large negative $V_G$ (start of FS). Hence, $V_T$ for FS is lower as compared to RS wherein a large positive $V_G$ at the start of the sweep aids electron capture at the $MoS_2/HfO_2$ interface[31,33] and by adsorbates[28,29] thereby increasing its $V_T$. This results in positive $\Delta V_T$ and clockwise hysteresis. However, the behavior of interface traps is different for an underlap device, especially for those in the underlap region without the gate. To understand the difference, electron energy band diagrams are shown (Figures 3b-e) in the gate-$HfO_2$-$MoS_2$ direction along two cutlines: (1) in the no underlap region (2) in the underlap region for four points on the $I_D$-$V_G$ ($V_D$ = 0.4 V, RT) data (Figure 3a) of the underlap FET. The fringing gate electric field experienced by the underlap region of the channel results in a lower effective gate voltage ($V'_G < V_G$) for the ungated channel (along cutline 2). At point 1, during FS, traps at the $MoS_2$/$HfO_2$ interface are unoccupied at low (negative) $V_G$ since the gate electric field direction is from $MoS_2$ towards the gate. This results in a low $V_T$. The band diagrams (Figure 3b) will be similar along the two cutlines with only a small difference in the extent of band bending- it will be slightly lower along cutline 2. As $V_G$ is increased to a higher positive value (point 2) during FS, the electric field direction reverses and electrons that are now attracted towards the $MoS_2$/$HfO_2$ interface get captured by the traps. Electron accumulation will be higher in the no underlap part of the channel with direct gate control. Hence, the Fermi energy ($E_F$) will be closer to the conduction band ($E_C$) along cutline 1 leading to higher trap occupancy compared to the underlap channel along cutline 2. The band diagrams for point 2 are illustrated in Figure 3c.

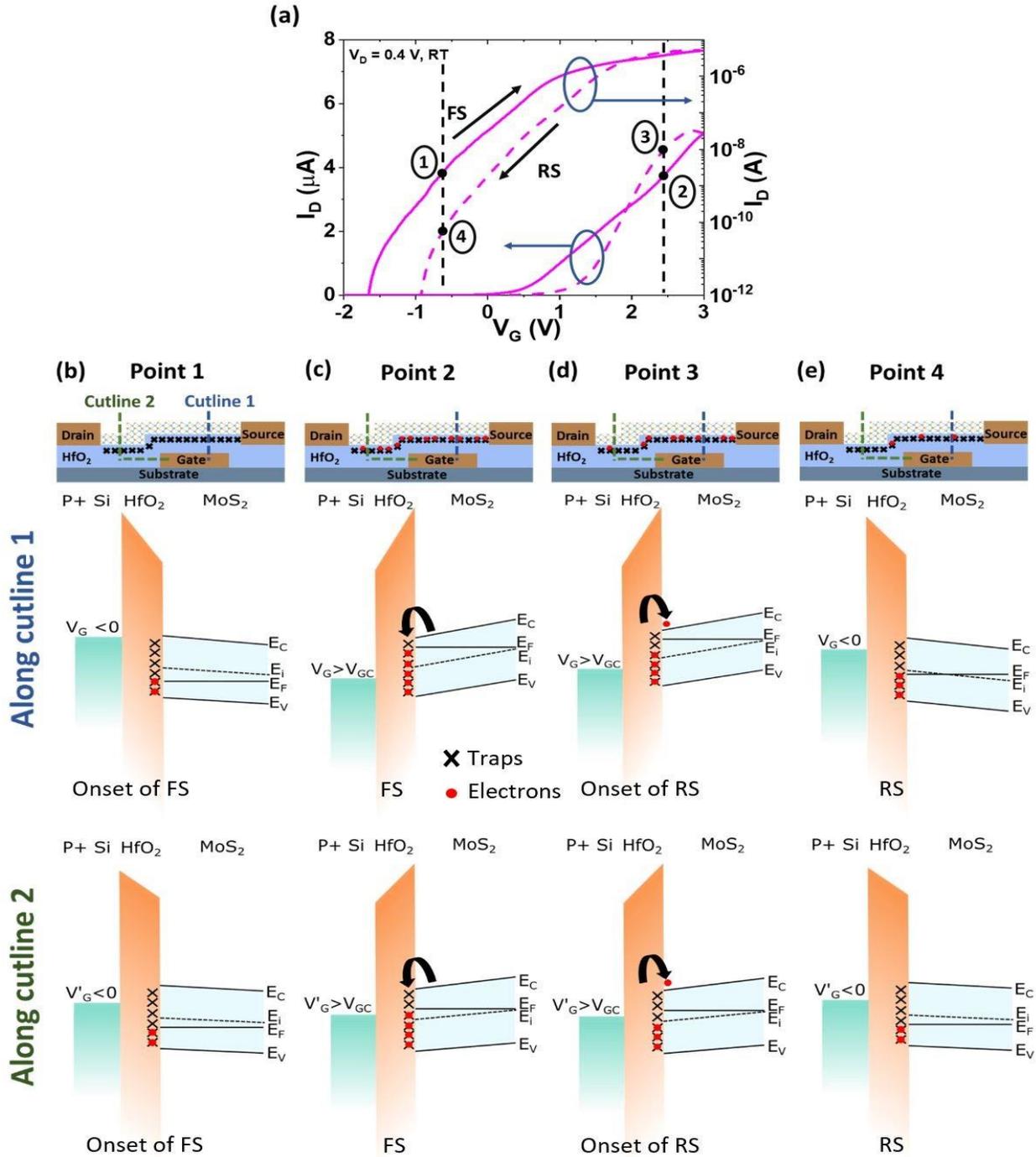

Figure 3: (a) Linear and log $I_D$-$V_G$ data at $V_D$ = 0.4 V for the underlap FET. Solid and dashed lines correspond to the forward and reverse sweeps, respectively. The 2D schematics and energy band diagrams along cutlines 1 and 2 [shown in (b)-(e)] depict the effect of gate voltage on traps at the $MoS_2$/$HfO_2$ interface (cross symbols) in the underlap device at (b) point 1, (c) point 2, (d) point 3 and (e) point 4 of the $I_D$-$V_G$ data in (a). $V_G$ represents the effective gate voltage experienced by the underlap channel and electrons are represented by red dots.

During RS, the electrons get de-trapped as $V_G$ is reduced to point 3 from +3 V. However, the de-trapping is more for the underlap channel along cutline 2 compared to cutline 1 due to the difference in gate fields. These de-trapped electrons counter the reduction in $I_D$ due to reduction in $V_G$, unlike the overlap device during reverse sweep and hence, the RS current (at point 3) is higher than the FS current (at point 2) giving rise to anti-clockwise hysteresis. As most of the traps near the underlap channel along cutline 2 de-trap with a further reduction in $V_G$ (point 4), a reduction in current is observed such that it is now similar to the overlap FET at the same RS $V_G$ (see Figure 2a), and the device's hysteresis changes from anticlockwise to clockwise (shown in Fig 3a). Figure 3d shows the band diagrams corresponding to point 3, where the $MoS_2$ $E_F$ corresponding to the underlap channel along cutline 2 is lower than along cutline 1 and hence, the occupied trap density is lower. When $V_G$ is reduced to point 4, the current is lower compared to point 1. This is due to higher threshold voltage during RS as there are still a few electrons trapped at the interface compared to FS.[28] It should also be noted that the occupancy of underlap interface traps, and hence $V_{GC}$ and anticlockwise hysteresis, can be significantly modulated by $V_D$. As $V_D$ is increased, the lateral drain electric field increases and prevents trapping of electrons at the underlap $MoS_2$/$HfO_2$ interface along cutline 2. This leads to a reduction in the anti-clockwise hysteresis window and a shift in crossover point ($V_{GC}$) to higher gate voltages (shown in Figure 2d).

To further investigate and quantify the trapping mechanisms in the MoS$_2$ FETs, measurements were carried out under vacuum and in ambient conditions at different $V_G$ sweep rates. Figures 4a-c show the transfer characteristics (log and linear plots of I$_D$-$V_G$) at $V_D$ = 0.5 V for the overlap and the underlap device in vacuum, and the overlap device in ambient conditions, respectively, for five different sweep rates. The appearance (absence) of clockwise hysteresis in ambient (vacuum) indicates that it is induced primarily by adsorbates.[28,29] For large negative $V_G$, adsorbates such as water and oxygen are repelled from the surface leading to electron release into the MoS$_2$ channel resulting in a low threshold voltage. However, for large positive $V_G$, these molecules are attracted towards the MoS$_2$ surface leading to electron capture from the channel. This increases the threshold voltage giving rise to clockwise hysteresis.[29] The clockwise hysteresis reduces with increasing sweep rate,[28,35–38] since it is limited by the trapping and de-trapping time constants of adsorbates. Therefore, the effective adsorbate trap density also reduces (shown in the inset of Fig. 4c), when estimated using:[37]

$$\Delta V_T^{ambient} - \Delta V_T^{vacuum} = \frac{qN_{adsorbates}}{C_{ox}} \qquad (1)$$

where $\Delta V_T^{ambient}$ ($\Delta V_T^{vacuum}$) is the hysteresis width of the overlap device measured in ambient (vacuum) conditions at room temperature, $q$ is unit electronic charge, N$_{adsorbates}$ is the effective density of adsorbate traps (number/cm$^2$) and C$_{ox}$ (=$\varepsilon_0\varepsilon_r/d$, where $\varepsilon_0$ is absolute permittivity, $\varepsilon_r$ is the relative permittivity of HfO$_2$ and $d$ is the thickness of HfO$_2$) is capacitance per unit area for the overlap device (1.39 $\mu$F/cm$^2$). It should be noted that the effective extracted adsorbate trap density for the underlap device is lower compared to the overlap device irrespective of sweep rate. The estimation of adsorbate trap density from the clockwise hysteresis widths of the overlap device under vacuum and in ambient indicates that the contribution of interface traps to clockwise hysteresis is relatively small. However, adsorbate traps cannot explain the hysteresis crossover in ambient and anti-clockwise hysteresis under vacuum in underlap devices as shown in Supporting Information S1 and Fig. 4b respectively. Unlike the reduction of clockwise hysteresis in ambient due to lower effective adsorbate trap density at higher sweep rates, the ambient anti-clockwise hysteresis in the underlap FET, indicated by $\Delta I_D/I_{D,FS}$ (shown in Fig. 4d), increases first at slower sweep rates before reducing at higher sweep rates. $V_{GC}$ decreases and then saturates with increasing sweep rate. These trends can be qualitatively explained by the combined effect of adsorbates and interface traps. At low sweep rates, the effect of adsorbates (CW hysteresis) dominates over interface trap induced ACW hysteresis, resulting in a large $V_{GC}$ or low $\Delta I_D/I_{D,FS}$. As the sweep rates increase, the effective adsorbate trap density (CW hysteresis) reduces, as shown in the inset of Fig. 4c, leading to larger ACW hysteresis, and therefore a peak in $\Delta I_D/I_{D,FS}$. With a further increase in sweep rate, the limited trapping and de-trapping time constants of the interface traps lead to a reduction in ACW hysteresis and hence, $\Delta I_D/I_{D,FS}$ reduces and $V_{GC}$ saturates.

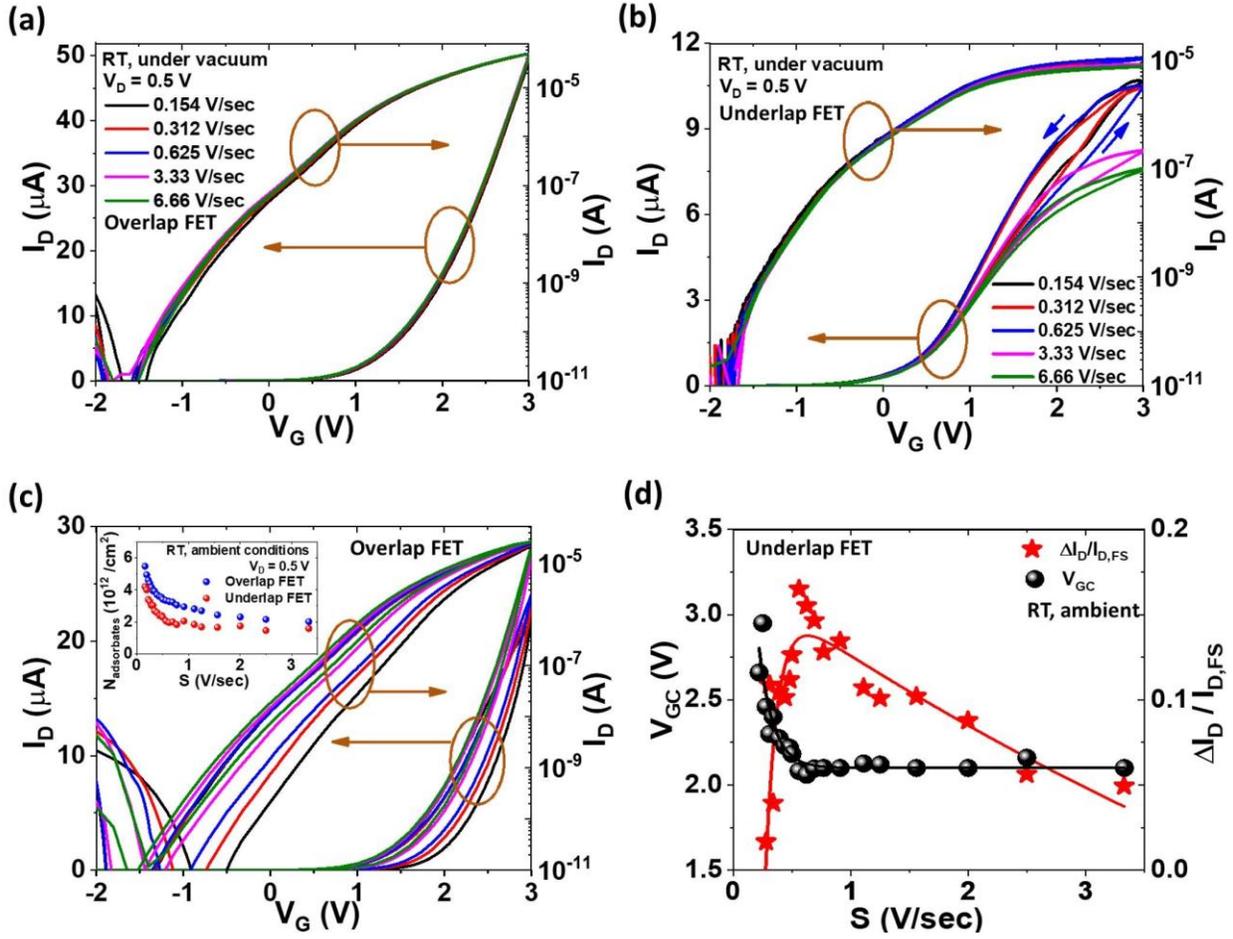

Figure 4: Transfer characteristics (linear and log plots) at $V_D = 0.5$ V for the (a) overlap and (b) underlap FETs in vacuum, and for the (c) overlap FET in ambient conditions for varying $V_G$ sweep rates at room temperature. The difference in hysteresis for both devices in vacuum and ambient conditions arises due to adsorbates giving rise to significant clockwise hysteresis in ambient conditions, which is absent when the devices are measured under vacuum. The inset in (c) shows the effective adsorbate trap density as a function of sweep rate for the overlap and underlap FETs. The presence of anti-clockwise hysteresis in the underlap FET even in vacuum reinforces the role of interface traps in its mechanism. (d) $V_{GC}$ and $\Delta I_D/I_{D,FS}$ (extracted at $V_G = 2.5$ V) for the underlap FET with varying sweep rates in ambient conditions.

Subthreshold characteristics of the FETs were also analyzed to extract the MoS$_2$/HfO$_2$ interface trap density that is responsible for the hysteresis crossover and anti-clockwise hysteresis in the underlap devices. The subthreshold current is given by,[39,40]

$$I_D = I_0 e^{\left(\frac{qV_{GS}}{nkT}\right)}\left[1 - e^{\left(\frac{-qmV_{DS}}{nkT}\right)}\right] = I_M\left[1 - e^{\left(\frac{-qmV_{DS}}{nkT}\right)}\right] \qquad (2)$$

where $I_0$ is the characteristic current that defines the current leaking through the channel, $V_{GS}$ and $V_{DS}$ are the gate-to-source and drain-to-source voltages respectively, same as $V_G$ and $V_D$ with source being grounded, $k$ is Boltzmann's constant and $T$ is the temperature. Plots of -ln[1-($I_D/I_M$)] vs $V_{DS}$ (shown in Supporting Information S2a and S2b for the overlap and underlap devices respectively) and subthreshold slope for both devices in ambient and vacuum conditions (shown in Supporting Information S2c) have been used to extract the values of $m/n$ and $n$ respectively. The detailed extraction method[39,40] has been described in Supporting Information S2. Table 1 summarizes the values of subthreshold slope, $n$, $m/n$ and $m$ for both devices at room temperature and a sweep rate of 0.625 V/sec.

Table 1: Summary of subthreshold parameters for overlap and underlap FETs at room temperature.

| Parameter | Overlap FET | Underlap FET |
|---|---|---|
| Subthreshold Slope (mV/dec) | 593 | 593 |
| $n$ | 9.92 | 9.92 |
| $m/n$ | 0.117 | 0.124 |
| $m$ | 1.16 | 1.23 |

$C_D=(m-1)C_{ox}$ (refer Supporting Information S2) gives a maximum depletion capacitance ($C_D$) of 0.22 $\mu$F/cm$^2$ for the overlap device. $C_D$ value for the underlap device is the same since both devices were fabricated on the same MoS$_2$ flake. This results in an oxide capacitance of 0.97 $\mu$F/cm$^2$ for the underlap device, which is lower than 1.39 $\mu$F/cm$^2$ of the overlap device. Lower net oxide capacitance in the underlap device is due to the lower, fringing gate capacitance in the underlap region. Based on these oxide capacitances, the effective interface trap densities for the overlap and the underlap device have been calculated as $7.62 \times 10^{13}$ /cm$^2$-eV and $5.29 \times 10^{13}$ /cm$^2$-eV respectively. The number of traps underneath the gates in the overlap and underlap devices will be significantly different, although the total number of interface traps will be nearly the same for both the devices with the MoS$_2$/HfO$_2$ interface being present throughout the channel length. The difference in effective interface trap densities is similar to the difference observed for the effective adsorbate trap densities due to the lower fringing gate capacitance of the underlap region. In fact, the ratios of effective adsorbate and interface trap densities are nearly the same- ~0.6 for adsorbates and ~0.7 for the interface traps at a sweep rate of 0.625 V/sec, for the two FETs.

Based on the interface trap and adsorbate mediated hysteresis mechanisms discussed above, the difference in the V$_T$s of the underlap and overlap FETs should be consistent with the interface and adsorbate trap densities. However, the threshold voltage difference obtained from these trap densities is significantly lower than the measured value. The threshold voltage difference ($\Delta V_T^{OU}$) can be computed using,

$$\Delta V_T^{OU} = V_{TO,\text{FS}} - V_{TU,\text{FS}} = \Delta V_{T,interface\ charge} + \Delta V_{T,depletion\ charge} + \Delta V_{T,adsorbates} \qquad (3)$$

where V$_{TO,FS}$ and V$_{TU,FS}$ are the threshold voltages of the overlap and underlap devices during FS respectively. $\Delta$V$_{T,interface\ charge}$, $\Delta$V$_{T,depletion\ charge}$ and $\Delta$V$_{T,adsorbates}$ are contributions to $\Delta V_T^{OU}$ of the V$_T$ differences due to interface traps, depletion charge and adsorbates respectively. Although the depletion charge is the same for both the devices, the difference in C$_{ox}$ values of the two devices will result in different threshold voltage contributions. Measurements in ambient resulted in a $\Delta V_T^{OU}$ value of 0.8 V whereas calculations resulted in 0.17 V arising primarily only due to $\Delta$V$_{T,depletion\ charge}$ ($\Delta$V$_{T,interface\ charge}$ and $\Delta$V$_{T,adsorbates}$ are negligible as C$_{it}$/C$_{ox}$ and C$_{adsorbates}$/C$_{ox}$ are nearly the same for both devices, where C$_{it}$ and C$_{adsorbates}$ are the capacitances associated with interface traps and adsorbates). This difference can be explained by the presence of fixed oxide charge in the HfO$_2$ gate dielectric and hence, the difference between the threshold voltages of the two devices can be modified as:

$$\Delta V_T^{OU} = V_{TO,\text{FS}} - V_{TU,\text{FS}}$$

$$= \Delta V_{T,\text{oxide charge}} + \Delta V_{T,\text{interface charge}} + V_{T,\text{depletion charge}} + \Delta V_{T,\text{adsorbates}} \quad (4)$$

where $\Delta V_{T,\text{oxide charge}}$ is the difference in $V_T$s of the two devices due to fixed oxide traps. Using the above equation and assuming the same oxide trap density in both devices, the oxide trap density is calculated as $1.36 \times 10^{13} / \text{cm}^2$.

To further understand the trap contributions to $\Delta V_T^{OU}$ and the anti-clockwise hysteresis behavior, transfer characteristics of the underlap and overlap devices were measured at higher temperatures (45 °C and 65 °C). Figures 5a and 5b show the transfer characteristics

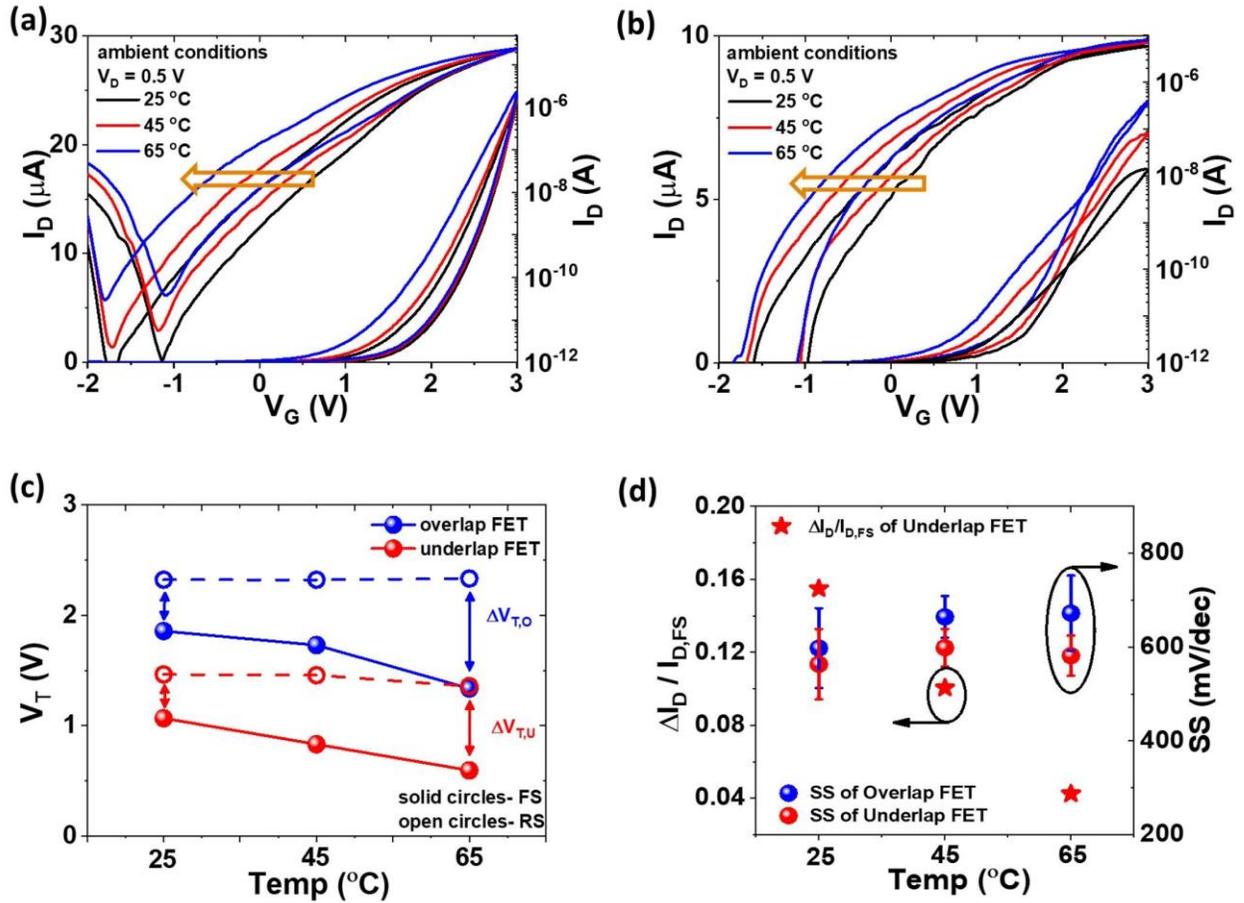

Figure 5: Transfer characteristics (linear and log plots) at $V_D$ = 0.5 V for the (a) overlap and (b) underlap FETs measured at three different temperatures in ambient conditions at a sweep rate of 0.625 V/sec. (c) Threshold voltage calculated for both forward and reverse sweeps ($\Delta V_{T,O} = V_{TO,FS} - V_{TO,RS}$, $\Delta V_{T,U} = V_{TU,FS} - V_{TU,RS}$) plotted for varying temperature, shows an increase in clockwise hysteresis at higher temperatures. (d) $\Delta I_D / I_{D,FS}$ (calculated at $V_G$ = 2.5 V) and SS for an underlap device plotted for varying temperature.

(log and linear plots) at three temperatures (25 ˚C (RT), 45 ˚C and 65 ˚C) in ambient at a sweep rate of 0.625 V/sec. As the temperature is increased, we observe (i) an increase in the clockwise hysteresis width of both devices since, (ii) the FS shifts more to the left (lower $V_G$) with the RS being nearly constant in both devices (shown in Fig. 5c), leading to, (iii) a decrease in the anti-clockwise hysteresis of the underlap device, as shown in Fig. 5d where $\Delta I_D/I_{D,FS}$ reduces with temperature. Adsorbates are likely not responsible for the left shift in FS at higher temperatures since previous studies have shown a reduction in clockwise hysteresis with increase in temperature due to loss of adsorbates at high temperatures.[18] Interface trap induced hysteresis, on the other hand, is expected to increase with temperature due to enhanced trapping and de-trapping of interface traps.[41] This increase in interface trap contribution, as also seen in the increase in the subthreshold slope with temperature in Fig. 5d, can explain the significant left shift of the FS transfer curves with temperature. However, the RS does not shift significantly for both devices and hence, not only the CW (ACW) hysteresis increases (decreases) but the difference between the reverse sweep $V_T$s of the underlap and overlap device also remains constant at all three temperatures. This constant difference in RS $V_T$s is indicative of a fixed oxide trap density of $1.36 \times 10^{13}$ /cm$^2$ (same as extracted from $\Delta V_T^{OU}$ at RT for forward sweep) which are thermally inactive even at a temperature of 65 ˚C. The oxide trap density, being lower than the interface trap density and inactive even at 65 ˚C explains the presence of clockwise hysteresis at all three temperatures. Table 2 summarizes the measurements carried out on both FETs for this study.

Table 2: Summary of the measurements done for overlap and underlap FETs.

| Measurement Conditions | Overlap | Underlap |
|---|---|---|
| $I_D$-$V_G$ (0.625 V/sec) for $V_D$ = 0.1-0.7 V | √ | √ |
| Ambient | √ | √ |
| Vacuum | √ | √ |
| Step size variation (0.154 V/sec to 3.33 V/sec) | √ | √ |
| Temperature variation (25, 45, 65 ˚C) | √ | √ |

## Conclusion

This work demonstrates a bottom gate underlap few-layer $MoS_2$ field effect transistor with room temperature clockwise hysteresis at lower gate voltages and anti-clockwise hysteresis at higher gate voltages. The ACW hysteresis width is tunable with the drain voltage, the underlap length and temperature, indicative of an interface trap induced hysteresis mechanism in the underlap FET at room temperature. By analysing electrical parameters such as hysteresis widths, subthreshold slopes and $V_T$ differences for both underlap and overlap devices in vacuum and ambient conditions, and at RT and high temperatures (45 °C and 65 °C), the contribution of each type of trap viz. adsorbates, traps at the $MoS_2/HfO_2$ interface and oxide traps, can be segregated. The underlap device architecture is highly promising for memory applications given a two fold advantage over previous studies: first the tunability of anti-clockwise hysteresis and its crossover with varying drain bias can enable read/write operations and second, the presence of hysteresis crossover at room temperature can lead to lower thermal budget and more reliable memories.

## Acknowledgement

Authors acknowledge Indian Institute of Technology Bombay Nanofabrication Facility (IITBNF) for the device fabrication and characterization. H.J. acknowledges Visvesvaraya PhD Scheme from Ministry of Electronics and Information Technology (Meity), Govt. of India. A.V. thanks IITB-Monash Research Academy for the doctoral fellowship and S.L. acknowledges the Department of Science and Technology (DST), Govt. of India through its SwarnaJayanti fellowship scheme (Grant number - DST/SJF/ETA-01/2016-17) for funding support.

## Supporting Information Available

It contains electrical characteristics of underlap device in ambient, and the extraction methodology for interface trap density.

# Supporting Information

# Electrically Tunable Room Temperature Hysteresis Crossover in Underlap MoS$_2$ FETs

Himani Jawa,[†] Abin Varghese,[†,‡] and Saurabh Lodha[*,†]

†Department of Electrical Engineering, IIT Bombay, Mumbai, Maharashtra 400076

‡Department of Materials Science and Engineering, Monash University, Clayton, Victoria, Australia 3800

E-mail: slodha@ee.iitb.ac.in

## S1: Transfer Characteristics of Underlap FET in Ambient Conditions

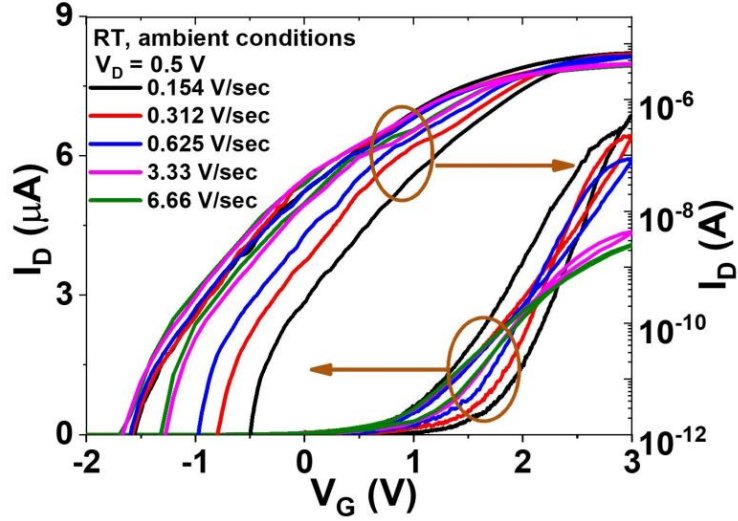

Figure S1: Transfer characteristics (linear and log plots) at $V_D$ = 0.5 V for the underlap FET in ambient conditions for different sweep rates at room temperature.

## S2: Extraction Methodology for Interface Trap Density

Transistor subthreshold current is given by[1,2],

$$I_D = I_0 \mathrm{e}^{\left(\frac{qV_{GS}}{nkT}\right)} \left[ 1 - \mathrm{e}^{\left(\frac{-qmV_{DS}}{nkT}\right)} \right] \qquad (1)$$

$$I_D = I_M \left[ 1 - \mathrm{e}^{\left(\frac{-qmV_{DS}}{nkT}\right)} \right] \qquad (2)$$

where

$$I_M = I_0 \mathrm{e}^{\left(\frac{qV_{GS}}{nkT}\right)} \qquad (3)$$

$$n = 1 + \frac{C_D + C_{it}}{C_{ox}} \qquad (4)$$

$$m = 1 + \frac{C_D}{C_{ox}} \qquad (5)$$

$I_0$ is the characteristic current that defines the current leaking through the channel, $V_{GS}$ and $V_{DS}$ are the gate-to-source and drain-to-source voltages, same as $V_G$ and $V_D$ with source grounded, $k$ is Boltzmann's constant, $T$ is the temperature, $C_D$ is the depletion capacitance per unit area and $C_{it}$ is the interface capacitance per unit area. Extracted subthreshold slope from the transfer characteristics for both the devices can be used to obtain $n$ using the equation,

$$n = \frac{Sq}{kT} \tag{6}$$

$m$ can then be calculated using,

$$m = \frac{kT}{q} \frac{d}{dV_{DS}} \left[ ln \left( 1 - \frac{I_D}{I_M} \right)^{-1} \right] \times n \tag{7}$$

Finally, the interface traps density can be calculated using the equation:

$$D_{it} = \frac{C_{it}}{q} = \frac{C_{ox}}{q} (n - m) \tag{8}$$

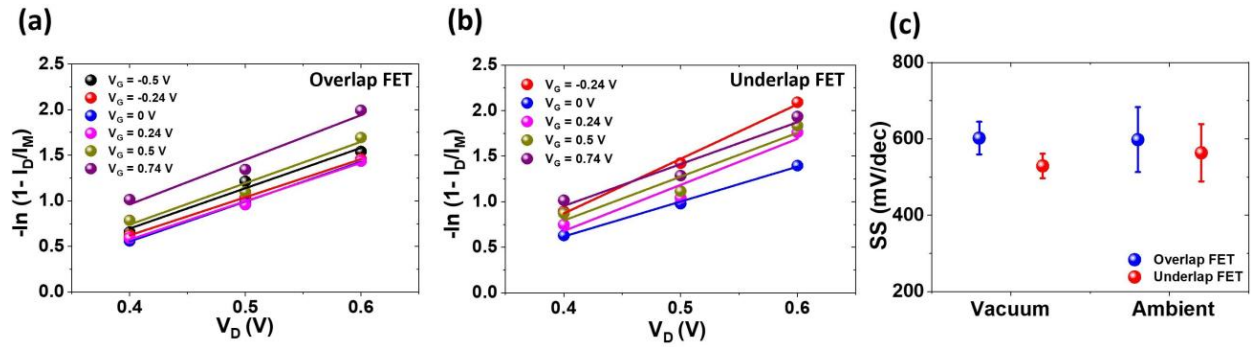

Figure S2: $ln(1-I_D/I_M)^{-1}$ vs $V_D$ for (a) overlap and (b) underlap FETs at different gate voltages in the subthreshold region. (c) Subthreshold slope plotted for both devices under vacuum and ambient conditions.